\documentclass[12pt,preprint]{aastex}

\shorttitle{Pressure Gradients and Rapid Migration}
\shortauthors{Haghighipour $\&$ Boss}

\begin{document}

\title{On Pressure Gradients and Rapid Migration of
       Solids in an Inhomogeneous Solar Nebula}

\author{ Nader Haghighipour and Alan P. Boss}
\affil{Department of Terrestrial Magnetism, Carnegie 
       Institution of Washington, \\ 
       5421 Broad Branch Road, Washington, DC 20015}

\email{nader@dtm.ciw.edu, boss@dtm.ciw.edu}

\begin{abstract}
We study the motions of small solids, ranging from micron-sized
dust grains to 100-m objects, in the vicinity of a local
density enhancement of an isothermal gaseous solar nebula. 
Being interested in possible application of the results to the
formation of clumps and spiral arms in a circumstellar disk,
we numerically integrate the equations of motion of such solids
and study their migration for different values
of their sizes and masses and also for different physical properties
of the gas, such as its density and temperature. We show that,
considering the drag force of the gas and also the gravitational
attraction of the nebula,
it is possible for solids, within a certain range of size and mass,
to migrate rapidly (i.e. within $\sim$1000 years) 
toward the location of a local maximum density 
where collisions and coagulation may result in an accelerated rate of
planetesimal formation.
\end{abstract}


\section{Introduction}

It is generally believed that planet formation
starts as a secondary process to star formation
by coalescence of small bodies in circumstellar disks.
With regard to our solar system,
two mechanisms have been proposed for the
formation of the giant planets in such a disk around our Sun; 
the widely accepted core accretion model \citep{Pol96} and
the disk instability scenario \citep{Bos00}.
It has recently been noted that a solar nebula massive enough
to possibly form giant planets via the core accretion model
is likely marginally-gravitationally unstable 
\citep{Pol96,Bos00,Inab02}. 
The alternative approach, namely the disk instability mechanism, 
however, implies that such an instability
could lead to rapid formation of gas giant planets.
It is, therefore, of great importance to study how the dynamics of 
small solids
will be affected in such an unstable environment, and what implications
there will be on the collision and coagulation process. 

In general, in a non-turbulent, freely rotating gaseous disk
at hydrostatic equilibrium,
there is a radial gradient associated with the gas pressure.  
This pressure gradient counteracts the gravitational attraction of the
central star and causes the gas molecules 
to have slightly different velocities than Keplerian circular.
When the pressure gradient is positive, the velocity of a gas molecule 
is greater than the local Keplerian velocity. 
A solid in the gas, in this case, feels an acceleration
by the gas along its orbit and, consequently, the increase in its
orbital angular momentum forces the solid 
to a larger orbit. In this case, we say that the solid
feels a ``tail wind.'' The opposite is true  when the pressure gradient
is negative. That is, a solid body will be subject to a ``head wind'' 
and will migrate toward smaller orbits.

One of the features of a rotating gravitationally unstable disk is the 
appearance of spiral arms or clumps where the density of the
medium is locally enhanced. In the vicinity of such density 
enhancements, the pressure of the gas may change radially and 
cause the particles in the disk
to migrate toward the location of the maximum gas density.
We are interested in studying the dynamics of solids that undergo
such migration and in exploring the possibility of applying the
results to the formation of planetesimals in a marginally-
gravitationally unstable disk. As the first stage of our project,
we present here the results of a systematic
study of the migration of solids subject to gas drag and the 
gravitational force of a circumstellar disk,
around the location of its maximum density. 
To focus attention on the dynamics of the
solids and its association with parameters such as
the temperature of the gas, the sizes of the objects, and also
the values of their densities, we consider a
hypothetical solar nebula with a circularly symmetric density
function.

Studies of the motions of solids in  gaseous mediums have been
presented by many authors. In a detailed analytical analysis in 1962,
Kiang studied the dynamical evolution of solids in elliptical orbits
subject to resistive forces proportional to arbitrary powers of their
relative velocities and their distances to the star. 
In his study, Kiang considered three cases of stationary, 
uniformly rotating, and also freely rotating gaseous mediums. 
However, he did not consider the pressure gradient of the gas.
It was  \citet{Whip64} who first 
mentioned that the rotation of the 
solar nebula deviates from Keplerian because of counterbalancing
the gravity of the Sun by the internal pressure of the gas which in turn
results in in/outward migration of small solids. In 1972, 
Whipple studied the dynamics of such solids in the solar nebula
where, following an approximation by \citet{ProbFas69},
he also included the resistive effect of the gas.
Whipple's work was subsequently expanded upon
and generalized by \citet{Weid77}
for a variety of model nebulae and different sizes of solids.

A comprehensive study of the effect of gas drag on the motions of
solid bodies can also be found in the classic work of \citet{Ada76}. 
In their paper, Adachi et al. studied
the motion of a solid on an elliptical orbit in a solar nebula whose 
density and temperature vary inversely with different powers of the
distance from the Sun. They also presented a detailed analysis of the form
of the gas drag for different relative velocities and 
relative sizes of solids, and also different values of the 
gas Reynolds number. 

Among the more recent studies of the dynamical evolution of solids
in a gaseous disk, one can name the work of \citet{Rano93} on
resonance capture of planetesimals subject to a drag force proportional 
to their relative velocities, as a barrier for the inward flow
of solids to the accretion zone of a planetary embryo in the solar
nebula, a paper by \citet{Sup00} on the formation of icy
planetesimals subject to a linear combination of Stokes and Epstein 
drags in an azimuthally symmetric, turbulent and thin solar nebula
with a polytropic equation of state, and also a paper by \citet{Iwas01} 
on the stability/instability of protoplanets subject to gas drag.
 
In this paper, we study the dynamics of solid bodies in 
an inhomogeneous gaseous disk. Our model nebula 
consists of a Sun-like star at its center and non-interacting
collisionless bodies scattered on its midplane.
We consider the effect of drag force  and also include the
gravitational attraction of the nebula.

The outline of this paper is as follows. 
Section 2 introduces the equations of motion and also the basic relations
concerning drag and the gravitational force of the gas. Section 3
defines the system of interest, and section 4 presents
the results of our numerical simulations. Section 5 concludes this
study by reviewing the results and discussing their applications.

\section{Basic Relations}

We consider a thin, isothermal and freely rotating 
gaseous disk with 
a Sun-like star at the center of its midplane.
A solid object in this medium, in addition to the gravitational
force of the central star, is subject to gas drag and also to
the gravitational attraction of the disk.
In a inertial coordinate system with its origin at the position of the star
and its axes on the midplane of the nebula,
the equation of motion of such a solid can be written as
\begin{equation}
{m_{\rm p}}\,{{\ddot{\it \bf r}}_{\rm p}}\,=\,
-\,GM{m_{\rm p}}
\biggl({{{\it \bf r}_{\rm p}}\over {r_{\rm p}^3}}\biggr)\,+\,
{{\it \bf F}_{\rm {drag}}}\,+\,{{\it \bf F}_{\rm {disk}}}\,,
\end{equation}
\vskip 2pt
\noindent
where $m_{\rm p}$ and ${\it \bf r}_{\rm p}$ 
represent the mass and the position
vector of the solid, $M$ is the mass of the central star
and $G$ is the gravitational constant. The quantities
${\it \bf F}_{\rm {drag}}$ and ${\it \bf F}_{\rm {disk}}$ in equation (1)
denote the gas drag and the gravitational force of the nebula, respectively. 
In the following, we discuss these two forces in detail.

\subsection{Gas Drag}

In general, the drag force of a gaseous medium with a density
$\rho_{\rm g}$ on a spherical body with radius
$a_{\rm p}$ is proportional to the square of its relative velocity, 
${\it \bf V}_{\rm rel}$, and is given by \citep{Lan59, Ada76}
\vskip 2pt
\begin{equation}
{{\it\bf F}_{\rm {drag}}}\,=\,-\,{1\over 2}{C_{\rm D}}\pi {a_{\rm p}^2}
{\rho_{\rm g}}{v_{\rm rel}}{{\it \bf V}_{\rm rel}}\,.
\end{equation}
\vskip 2pt
\noindent
In this equation, ${v_{\rm rel}}=|{{\it \bf V}_{\rm rel}}|$ and $C_{\rm D}$,
the drag coefficient, is a dimensionless constant that 
depends on the gas Reynolds number, the ratio of the relative velocity 
of the solid to
the speed of sound in the medium (the Mach number), and also the relative 
size of the solid compared to the mean free path of the gas molecules 
(the Knudsen number). For a 
detail analysis of the drag coefficient
we refer the reader to \citet{Ada76}. 

For the cases where
the mean free path of the gas molecules is smaller than the size
of the object, we have \citep{Whip72, Weid77}
\vskip 2pt
\begin{equation}
{C_{\rm D}}\,\simeq\,\cases{
24\,{R{\rm e}^{-1}} &if$\quad R{\rm e}<1$ (Stokes' drag) \cr
24\,{R{\rm e}^{-0.6}}&if$\quad1<R{\rm e}<800$\cr
0.44&if$\quad R{\rm e}>800$,\cr}
\end{equation}
\vskip 2pt
\noindent
where $R{\rm e}$ is the gas Reynolds number. 
For a gas with a viscosity $\nu$, 
\begin{equation}
R{\rm e}\,=\,{2\over \nu}\,{\rho_{\rm g}}\,{a_{\rm p}}\,{v_{\rm rel}}\,,
\end{equation}
\vskip 2pt
\noindent
and
\begin{equation}
\nu\,=\,{1\over 3}\,\Bigl({{{m_0}{{\bar v}_{\rm th}}}\over \sigma}\Bigr)\,,
\end{equation}
\vskip 2pt
\noindent
where $m_0$ and ${\bar v}_{\rm th}$ represent the
mass and the mean thermal velocity of the gas molecules and
$\sigma$ is their collisional cross section 
\citep{Ada76, Weid77}. 

For particles moving much slower than the gas mean thermal 
velocity and with sizes
smaller than the mean free path of the gas molecules, 
${{\it \bf F}_{\rm {drag}}}$ can
be approximately written as
\begin{equation}
{{\it \bf F}_{\rm {drag}}}\,=\,-\,{4\over 3}\,\pi\,
{\rho_{\rm g}}\,{a_{\rm p}^2}\,
{v_{\rm th}}\,{{\it \bf V}_{\rm rel}}\,.
\end{equation}
\vskip 2pt
\noindent
Equation (6) is known as Epstein drag \citep{Ken38, Eps24}.

As shown by equations (2) and (6), the resistive effect of the
gas is directly proportional to the size of solids.
Anticipating numerically integrating equation (1) for
different values of $a_{\rm p}$, we
follow \citet{Sup00} and 
combine equations (2) and (6) by introducing 
$f={a_{\rm p}}/({a_{\rm p}}+\ell)$
where $\ell$ is the mean free path of the gas molecules.
We now write
\vskip 1pt
\begin{equation}
{{\it \bf F}_{\rm {drag}}}\,=\,-\,
{4\over 3}\pi{a_{\rm p}^2}\,{\rho_{\rm g}}\,
\Bigl[(1-f){{\bar v}_{\rm th}}\,+\,{3\over 8}f{C_{\rm D}}{v_{\rm rel}}\Bigr]\,
{{\it \bf V}_{\rm rel}}\,,
\end{equation}
\vskip 2pt
\noindent
which is particularly useful for transitional cases where
the size of solids and the mean free path of the gas
molecules are comparable. In equation (7),
${{\it \bf V}_{\rm rel}}={{\it \bf V}_{\rm p}}-{{\it \bf V}_{\rm g}}$ 
where ${{\it \bf V}_{\rm p}}$ is the velocity of a solid and
${\it \bf V}_{\rm g}$, the velocity of the gas
at the location of the solid, has a magnitude
given by
\vskip 2pt
\begin{equation}
{v_{\rm g}^2}\,=\,{{GM}\over {r_{\rm p}}}\,+\,
{{r_{\rm p}}\over {{\rho_{\rm g}}({{\it \bf r}_{\rm p}})}}\,
{\Biggl({{d{P_{\rm g}}}\over {dr}}\Biggr)_{{\it \bf r}=
{{\it \bf r}_{\rm p}}}}\,.
\end{equation}
\noindent
In this equation, $P_{\rm g}$ is the pressure of the gas. As shown
in equation (8), the velocity of the gas slightly differs
from its Keplerian circular value (first term of the right hand side)
due to the pressure gradient.

\subsection{Gravitational Potential of the Disk}

As mentioned earlier in this section, the model nebula considered 
here is a gaseous disk with a radius much larger than its thickness.
Although thin, this nebula has a considerable thickness compared
to the size of the solids of our interest. Therefore,
in addition to the drag forces and the gravitational attraction of 
the central star,
objects in this nebula feel a force associated with the gravitational
potential of the disk. In this section, we present a brief analytical
analysis of the calculation of the gravitational force of the
above-mentioned nebula. Detailed analysis is in preparation for 
publication elsewhere.

As seen by a solid in the nebula, the gaseous disk surrounding
the central star resembles a thin cylinder with a considerably large
radius. The gravitational force of the nebula on a solid
is equal to the gradient of the gravitational potential
of this cylinder at the location of the solid. This potential function,
denoted by $\phi({\it \bf R})$, satisfies Poisson's equation. That is,
\begin{equation}
{\nabla^2}\phi({\it \bf R})\,=\,4\pi G {\rho_{\rm g}}({\it \bf R})\,.
\end{equation}
\vskip 2pt
\noindent
In this equation, $\it {\bf R}$ is the solid's three dimensional
position vector in a general coordinate system with an origin 
at the position of the central star. 
For the system of interest in this paper, it is possible to
reduce the partial differential equation (9) into a more manageable form.
As mentioned in section 1, we consider ${\rho_{\rm g}}$ to be
circularly (azimuthally) symmetric. 
We also assume that the motions of solids 
are restricted to the midplane of the nebula.
Since the thickness of the disk
is much larger than the size of the solids, 
the latter assumption implies that the change of the gravitational
potential of the gas along the diameter of a solid perpendicular to 
the midplane of the nebula can be considered
negligible. Combined with the azimuthal 
symmetry of ${\rho_{\rm g}}$, in a cylindrical coordinate system
with its origin on the central star and its 
plane-polar components measured on the midplane
of the nebula,
the simplifying assumption above allows us to write equation (9) as
\begin{equation}
{1\over r}\,{d\over {dr}}\,\Bigl(r\,{{d\phi}\over {dr}}\Bigr)
\,=\,4\pi G\,{\rho_{\rm g}}(r)\,,
\end{equation}
\vskip 2pt
\noindent
where $r$ is the plane-polar component of $\it {\bf R}$.
The gravitational force associated with potential 
$\phi({\it \bf R})$ is equal to $-m{\it \bf \nabla}\phi({\it \bf R})$
and for a particle restricted to move on the midplane of the nebula
is given by
\vskip 2pt
\begin{equation}
{{\it \bf F}_{\rm {disk}}}({\it \bf r})\,=
\,-\,{{4\pi}\over{r^2}}\,G\,m\,
\biggl[\int {\rho_{\rm g}} (r') r' dr'\,-\,c\,\biggr]\,{\it \bf r}\,,
\end{equation}
\vskip 20pt
\noindent
where $c$ is the constant of
integration whose value is calculated from the dynamical
properties of the system.

\subsection{Equation of Motion}

The equation of motion of a solid in this
study is given by equation (1). For the purpose of numerical
integrations, it is more convenient
to write this equation in a dimensionless form. Introducing the
quantities $r_0$ and $t_0$ which carry the dimensions of length
and time, respectively, equation (1) can be written as
\vskip 10pt
\begin{equation}
{\ddot{\it \bf{\hat r}}}\,=
\,-\,{\hat k}\biggl({{\it \bf{\hat r}}\over {{\hat r}^3}}\biggr)
\,+\,{{\it \bf{\hat F}}_{\rm {drag}}}\,+\,{{\it \bf{\hat F}}_{\rm {disk}}}\,.
\end{equation}
\vskip 2pt
\noindent
where  ${{\it \bf r}_{\rm p}}={r_0}{\it \bf {\hat r}},\, t={t_0}{\hat t},\,  
{\hat k}=GM{t_0^2}/{r_0^3}$ and a general dimensionless 
force $\it \bf {\hat F}$ is equal to 
${t_0^2}{\it \bf F}/{m_{\rm p}}{r_0}$. We choose
$r_0$ and $t_0$ such that ${\hat k}=1$. For 
a solid restricted to move on the midplane of
the nebula, equation (12), in a dimensionless form
and in a plane-polar coordinate system with axes on the
midplane of the disk, is written as
\begin{eqnarray}
&\!\!\!\!\!\!\!\!\!\!\!\!\!{P_r}=&\!\!\!{\dot r},\\
&\!\!\!\!\!\!\!\!\!\!\!\!\!{P_\theta}=&
\!\!\!{r^2}\,{\dot \theta},\\
&\!\!\!\!\!\!\!\!\!\!\!\!\!{{\dot P}_r}=&
\!\!\!\!{1\over {r^3}}{P_\theta^2}-{1\over {r^2}}-
{1\over r}\biggl[4\pi\int {{\tilde \rho}_{\rm g}}(r')r'dr'-c\biggr]
-{4\over 3}\pi{a_{\rm p}^2}{{\hat \rho}_{\rm g}}(r){P_r}
\Bigl[(1-f){{\bar v}_{\rm th}}+{3\over 8}f{C_{\rm D}}
{v_{\rm rel}}\Bigr],\\
&\!\!\!\!\!\!\!\!\!\!\!\!\!{{\dot P}_\theta}=&
\!\!\!\!-{4\over 3}\pi r{a_{\rm p}^2}
{{\hat \rho}_{\rm g}}(r){\bigl({v_{\rm rel}^2}-{P_r^2}\bigr)^{1/2}}
\Bigl[(1-f){{\bar v}_{\rm th}}+{3\over 8}f{C_{\rm D}}{v_{\rm rel}}\Bigr],
\end{eqnarray}
\vskip 2pt
\noindent
where $P_r$ and $P_\theta$ are, respectively,
the radial and the angular momenta of the solid,
${{\hat \rho}_{\rm g}}\,=\,{r_0^3}{\rho_{\rm g}}(r)/{m_{\rm p}}$,
and 
${{\tilde \rho}_{\rm g}}\,=\,{r_0^3}{\rho_{\rm g}}(r)/M$. 
In equations (13) to (16), the hat signs have been dropped
for simplicity.

\section {The Physical Model}
 
We consider a gaseous nebula of pure molecular hydrogen with
a uniform temperature $T$ and a density given by
\begin{equation}
{\rho_{\rm g}}(r)\,=\,{\rho_0}\,e^{-\alpha{(r-{r_{\rm m}})^2}}\,.
\end{equation}
\vskip 2pt
\noindent
In equation (17), $\rho_0$ is the magnitude of the local 
maximum of the density and $r_{\rm m}$ denotes its radial
position. The coefficient $\alpha$ in this equation 
is a positive constant. Figure 1 shows ${\rho_{\rm g}}(r)$ for
$\alpha = 0.5$ and 1.
It is necessary to  emphasize that the choice of density,
as given by equation (17), has been made, solely, 
to focus attention on the  effects 
of the pressure gradient, gas drag  
and also the gravitational force of the disk on migration of solids and their 
dynamics. In a more realistic system, the density of the gas will
have a much more complicated form. Extension of this work to such
cases is the subject of up-coming articles.\notetoeditor{Fig. 1 
on the same page as the first paragraph of Sec. 3, in print.}

As mentioned earlier, we would like to study the dynamics of a solid
in a gaseous disk with a density given by equation (17)
by numerically simulating its motion given by
equations (13) to (16). These equations
require us to write certain quantities such as the pressure of the gas,
its mean thermal velocity, the mean free path of its molecules
and also the relative velocity of the solid,
in terms of the gas density $\rho_{\rm g}$. To do so,
we assume that our model nebula obeys 
the equation of state of an ideal gas, ${P_{\rm g}}=n{k_{\rm B}}T$, where 
$n$ is the gas number density and $k_{\rm B}$ is the Boltzmann's constant.
The mean thermal velocity of the gas molecules is, therefore,
given by
\begin{equation}
{{\bar v}_{\rm th}^2}\,=\,{{8{k_{\rm B}}T}\over{\pi{m_0}}}\,.
\end{equation}
\vskip 2pt
\noindent
Substituting for $T$
from the ideal gas law in equation (18), the pressure of the gas
will be equal to 
\begin{equation}
{P_{\rm g}}\,=\,{1\over 8}\,\pi\,{{\bar v}_{\rm th}^2}\,{\rho_{\rm g}}(r)\,.
\end{equation}
\vskip 2pt
\noindent
From this equation and in a dimensionless form, 
the relative velocity of a solid can be
written as
\vskip 2pt
\begin{equation}
{v_{\rm rel}^2}\,=\,{P_r^2}\,+\,
{\Biggl[{{P_\theta}\over r}\,-\,
{\biggl({1\over r}+{{\pi r {{\bar v}_{\rm th}^2}}\over {8{\rho_{\rm g}}(r)}}
{{d{\rho_{\rm g}}(r)}\over {dr}}\biggr)^{1/2}}\Biggr]^2}\,.
\end{equation}
\vskip 6pt

The mean free path of the gas molecules can also be written in
terms of $\rho_{\rm g}$. Assuming a diameter of $a_0$ for the gas
molecules,  $\ell=1/(\pi{a_0^2}n)$. From the ideal gas law and also
from equation (18), one can write
\begin{equation}
\ell\,=\,{{m_0}\over {\pi{a_0^2}{\rho_{\rm g}}(r)}}\,.
\end{equation}
\vskip 3pt
\noindent
For molecular hydrogen, ${a_0}=1.5\times{10^{-8}}$ cm. The
mean free path of the molecules of our model nebula can, therefore,
be written as, 
$\ell$ (cm)$={{4.72\times{10^{-9}}}/ {{\rho_{\rm g}}(r)}}$ 
(g cm$^{-3}$).

The gravitational attraction of the disk as given by equation (11)
requires us to calculate the integral 
$\int {{\rho}_{\rm g}}(r) r dr$ and also the constant of integration $c$.
Using equation (17), the integral of
equation (11) will be equal to
\vskip 1pt
\begin{equation}
\int {{\tilde \rho}_{\rm g}} (r') r' dr'\,=\,{1\over 2}\,{\alpha^{-1/2}}\,
{\tilde {\rho_0}}\,
\biggl\{{r_{\rm m}}{\pi^{1/2}} 
{\rm Erf}\bigl[{\alpha^{1/2}}(r-{r_{\rm m}})\bigr]\,-\,
{\alpha^{-1/2}}\,{e^{-\alpha{(r-{r_{\rm m}})^2}}}\biggr\}\,,
\end{equation}
\vskip 3pt
\noindent
where ${{\tilde \rho}_0}={r_0^3}{\rho_0}/M$ and Erf$(r)$ is the 
error function.
As mentioned in section 2.2, the constant of integration $c$ 
depends on the dynamical properties of the system. In
our physical model with a density function given by equation (17),
the radial migration of solids due to the pressure gradient terminates at
$r={r_{\rm m}}$. That is, a solid object
at position $r_{\rm m}$ will stay with the gas and, consequently,
its relative velocity will become zero. That implies, 
at the location of the maximum density of the
gas, ${{\it \bf F}_{\rm {drag}}}=0$. Also, because we assumed that the disk
is rotating freely, a solid at $r={r_{\rm m}}$ will rotate the central star
in a circular orbit. That means, at this position, ${\dot r}=0$.
Applying these two conditions to equation (22), we have
\vskip 1pt
\begin{equation}
c\,=\,{1\over {r_{\rm m}}}\,-\,{r_{\rm m}^2}
{{{\dot \theta}^2}_{(r={r_{\rm m}})}}\,-\,
{{2\pi {{\tilde \rho}_0}}\over \alpha}\,,
\end{equation}
\vskip 2pt
\noindent
where all quantities have been written in a dimensionless form.
\vskip 25pt
\section {Numerical Simulations}
\vskip 5pt

We consider a gas density given by equation (17) with a peak
equal to ${10^{-9}}$ g cm$^{-3}$ at ${r_{\rm m}}=1$ AU. The value of
$\alpha$ is considered as a parameter in our simulations.
As mentioned in section 3, we choose our units such that 
${\hat k}=1$. Therefore, the quantities $r_0$ and $t_0$ are
related as ${t_0^2}={r_0^3}/GM$. Introducing a new
variable ${T_0}=2\pi {t_0}$, we have
\vskip 1pt
\begin{equation}
{T_0}\,=\,2\pi {\Bigl({{r_0^3}\over {GM}}\Bigr)^{1/2}}\,.
\end{equation}
\vskip 2pt
\noindent
Equation (24) implies that $T_0$ can be considered as the
period of an object rotating uniformly around the star $M$
on a circular path with radius $r_0$. In our physical
model, an object at $r={r_{\rm m}}$ has such a uniform circular
motion. We set ${r_0}={r_{\rm m}}=1$ AU and, 
therefore, from equation (24), 
${t_0}\simeq 5.03 \times {10^6}$ sec$\,\simeq 0.16$ years.
Also in this case, ${\dot\theta}({r_{\rm m}}=1$ AU) = 1 and
$c\simeq-0.0106\,{\alpha^{-1}}$.

We numerically integrated equations (13) to (16) for different values
of the gas temperature, solids' radii and densities, and also 
different values of $\alpha$. Figure 2
shows the radial migration of solids ranging from micron-sized
particles to 100 m objects from 2 AU and 0.25 AU to the location 
of the peak of the density of Figure 1 with $\alpha =1$.
The densities of the objects are equal to 2 g cm$^{-3}$.
As expected, small particles spend more time
with the gas and take a 
longer time to migrate in/outward. As the
sizes of the particles increase, while their densities stay constant,
the rate of radial migration also increases. Figure 3 shows a
comparison of the times of migration 
for four different values of the initial orbital radii of the solids of
Figure 2. One can see from this figure that for the above-mentioned physical
properties of the gas and solids, the graphs of the 10 cm
and 1 m objects show the most rapid radial migration, comparable
to the timescale of the growth of non-axisymmetries in 
disk instability models.
Increasing the radius of the solid to 10 m or higher, one observes that
the rate of migration decreases again. 
\notetoeditor{Figures 2 and 3 preferably on two pages adjacent to 
one another with the paragraph starting with ``We numerically ....''
on one of those pages.}

Figure 3 also shows that the rates of inward and outward 
migrations for equal distances
at both sides of $r=1$ AU are different. The inward migrations from 1.75, 1.5
and 1.25 AU occur more rapidly than the outward migrations from
0.25, 0.5 and 0.75 AU. To explain these differences, we have
plotted the relative velocity of a particle (Eq. [20]) 
during its migration. Figure 4 shows the magnitude of 
${\it \bf V}_{\rm {rel}}$ for a 1 m-sized object with a density equal 
to 2 g cm$^{-3}$ migrating from $r=$1.75 or 0.25 AU to $r=1$ AU. 
The temperature of the gas is 1000 K. As shown here, the magnitude
of the relative velocity of the object when approaching $r=1$ AU 
from a larger distance is greater than its relative 
velocity when migrating out. One can explain this, qualitatively,
by studying equation (20) in more details. 
\notetoeditor{Figure 4 on the same page as this paragraph in print.}
From Figure 1, the quantity $d{\rho_{\rm g}}(r)/dr$,
the contribution of the pressure gradient, is negative
for distances larger than 1 AU and positive for smaller distances.
When migrating in/outward, the absolute value of the quantity above 
decreases while its sign stays the same. Since
the contribution of the pressure gradient in case of an
outward migration is positive, the 
second term inside the square brackets of equation (20)
will have larger values for outward migrations resulting in
smaller values for the second term of equation (20) and,
consequently, smaller relative velocities in these cases.
From equation (2), this implies that resistive effect of the headwind
felt by the solid during its inward migrating is larger than the
effect of the tailwind that causes
its outward motion. In other words, the inward migration of the
solid is accomplished faster.

The rate of radial migration varies also with the density of
solids. Figure 5 shows the migrations of solids with radii from
1 cm to 10 m for three different values of the solid's density.
The physical properties of the gas in this figure are identical to
Figure 1. As shown in Figure 5, for $a_{\rm p}$=1 and 10 cm, 
the rate of radial migration increases by increasing the solid's
density. The reason for this can be found in the contribution of
the drag force (Eq. [7]) to the change of the radial momentum 
of the solid given by equation (15). We recall that in the latter 
equation, ${\hat \rho}_{\rm g}$ is a dimensionless quantity whose numerical 
value is equal to ${\rho_{\rm g}}{r_0^3}/{m_{\rm p}}$. 
For a constant value of $a_{\rm p}$, increasing the density
of the object will result in increasing its mass and, consequently,
decreasing the dimensionless density ${\hat \rho}_{\rm g}$. 
As a result, the
last term of equation (15) will have a smaller absolute value and,
therefore, the radial momentum 
of the object will have a larger
rate. That is, its radial migration will be faster.\notetoeditor{Figure 
5 on the same page as this paragraph in print.}

The above-mentioned feature does not sustain for larger
values of the solids radii. When $a_{\rm p}\ge 1$ m, 
both the inward and the outward
migrations slow down when the densities of solids are increased. 
This can also be attributed to the behavior of the last term of 
equation (15) for these values of $a_{\rm p}$. Recall that from 
equation (21), for the range of $r$ considered here
(i.e., $0.25\leq r \leq 2$) , the maximum
value of the mean free path of the gas molecules is approximately
12.8 cm. That means, for the values of the solids radii larger than
1 m, the coefficient $f$ is approximately equal to one. The
contribution of the drag force of the gas to the time variation
of the radial momentum
of the solids given by the last term of equation (15), in this case, 
will be
proportional to $a_{\rm p}\,,\,{a_{\rm p}^{1.4}}{\rho_{\rm g}^{0.4}}$,
and ${a_{\rm p}^2}{\rho_{\rm g}}$ for different values of the drag
coefficient $C_D$ given by equation (3). It is evident that for
meter-sized and larger objects, the last term of equation (15) will have
larger absolute values resulting in smaller values for ${\dot P}_r$ and,
consequently, slower radial migration.

Numerical integrations have also been carried out for different values of the
temperature of the gas. The effect of a change in temperature appears in
the thermal velocity of the gas which in turn results in changes in the
value of the gas viscosity and eventually 
its drag coefficient, and also
affects the pressure gradient and the relative velocity of
the solids. Figure 6 shows the rate of migration of a 10 cm-sized
solid with a density of 2 g cm$^{-3}$ for two values of 
$\alpha=0.5$ and 1. As shown here, by increasing the temperature,
the time of migration toward the location of the maximum density
increases. This can be attributed to the fact that 
an increase in the temperature of the gas will
increase the mean thermal velocity of its molecules and,
therefore, from equation (20), will result in smaller values of
the relative velocity of the solid. 
The drag force of the gas, in this case, will have a smaller
magnitude (see eq. [2]) implying that 
when migrating inward at a higher temperature, the drag force
will need more time to slow down the object and when migrating outward,
the import of angular momentum to the particle via a tailwind
generated by the gas drag will be smaller.
Figure 7 shows a comparison between the magnitude of the relative
velocity of a 10 cm particle migrating inward from 2 AU and outward
from 0.25 AU at 1000 K and 300 K. One can see from this figure that
for both cases of inward and 
outward migrations, the magnitude
of the relative velocity is larger for a smaller temperature.

Although the above-mentioned
features are observed in the motion of a solid in both cases
of $\alpha=0.5$ and 1, a
comparison between the graphs of these two values of $\alpha$ indicates
that for each value of the temperature, the time of migration is
longer for smaller values of $\alpha$, a result that is 
evident from the density distribution shown in
Figure 1; smaller values of $\alpha$ corresponds to smaller
gas pressure gradients.
\notetoeditor{If possible, in print, figures 6 and 7 on a page 
before section 5.}
\vskip 35pt
\section {Summary and Discussion}
\vskip 5pt

We have studied motions of small solids in the vicinity of a
density enhancement in a freely rotating gaseous disk. We assume that
the gas is isothermal and ideal. In this case, an enhancement 
in the gas density corresponds to a maximum in its pressure. As expected, 
because of the pressure gradient associated with the radial change
of the gas density, solids on both sides of the location of the maximum 
pressure (or density) undergo in/outward migrations. We
have studied such migrations in a 
model solar nebula where solids, in addition to the 
gravitational attraction of the central star, are subject to
gas drag and to the gravitational force of the nebular disk.

In general, the rate of the migration of a solid in a gaseous medium due to
the pressure gradient varies with the solid's mass and also with
the physical properties of the gas such as its density and temperature.
As shown in section 2, changes in the gas temperature and density will
affect the mean thermal velocity and also  
the mean free path of the gas molecules. Such changes 
show their effects in the drag force of the gas through its Reynolds 
number and also through the relative velocity of the solid and
result in different times of migration.
An analytical study of the general dependence of the rate of migration
on the physical properties of the solids and the nebular gas is currently
under way.

As mentioned earlier, our motivation for initiating this study was to 
seek the possibility of the application of the results to the formation 
of planetesimals in marginally-gravitationally unstable
disk models. As shown by \citet{Bos00},
the disk instability scenario suggests rapid formation of giant 
gaseous protoplanets followed by
sedimentation of small solids at the location of spiral
arms and clumps of a gravitationally unstable disk all
in about one thousand years. In our study, we considered
a simple model of the solar nebula in order to focus our attention
solely on the time of migration of solids and their variations with
physical parameters of the system. Our results indicate that it is, 
indeed, possible for solids within certain ranges of size and 
density to migrate quite rapidly to the locations of maximum values
of the gas density. For the model studied here, solids with densities
of a few g cm$^{-3}$ and with radii ranging from several centimeters
to a few meters, migrate a radial distance of 1 AU during a time
($\sim$10$^3$ years)
comparable with the giant planet formation timescale implied by the 
disk instability model. 

Regardless of whether disk instability can form gas giant planets,
the likelihood that the solar nebula was marginally-gravitationally
unstable implies that the processes studied here may have enhanced
the growth rates of solid planetesimals.

At the end we would like to mention that the calculations in this study
have been done for two dimensional motions and in a solar nebula 
that may not fully reflect the properties of a more realistic 
environment. In this study, we focused on the motion of solids 
as isolated objects without including their 
mutual interactions. 
To obtain a better understanding of the dynamics of solids 
and the times of their migrations, it is necessary to extend such
an analysis to a three dimensional case and to allow for interactions
between the objects. It is also important to consider 
temperature and density distributions for the nebular gas that
portray the physical properties of a gravitationally unstable
disk in a more realistic way. Such considerations are currently
under way.

\acknowledgments

This work is partially supported by the NASA Origins of the Solar 
System Program  under Grant NAG5-10547 and by the NASA Astrobiology
Institute under Grant NCC2-1056.

\clearpage

\begin{figure}
\plotone{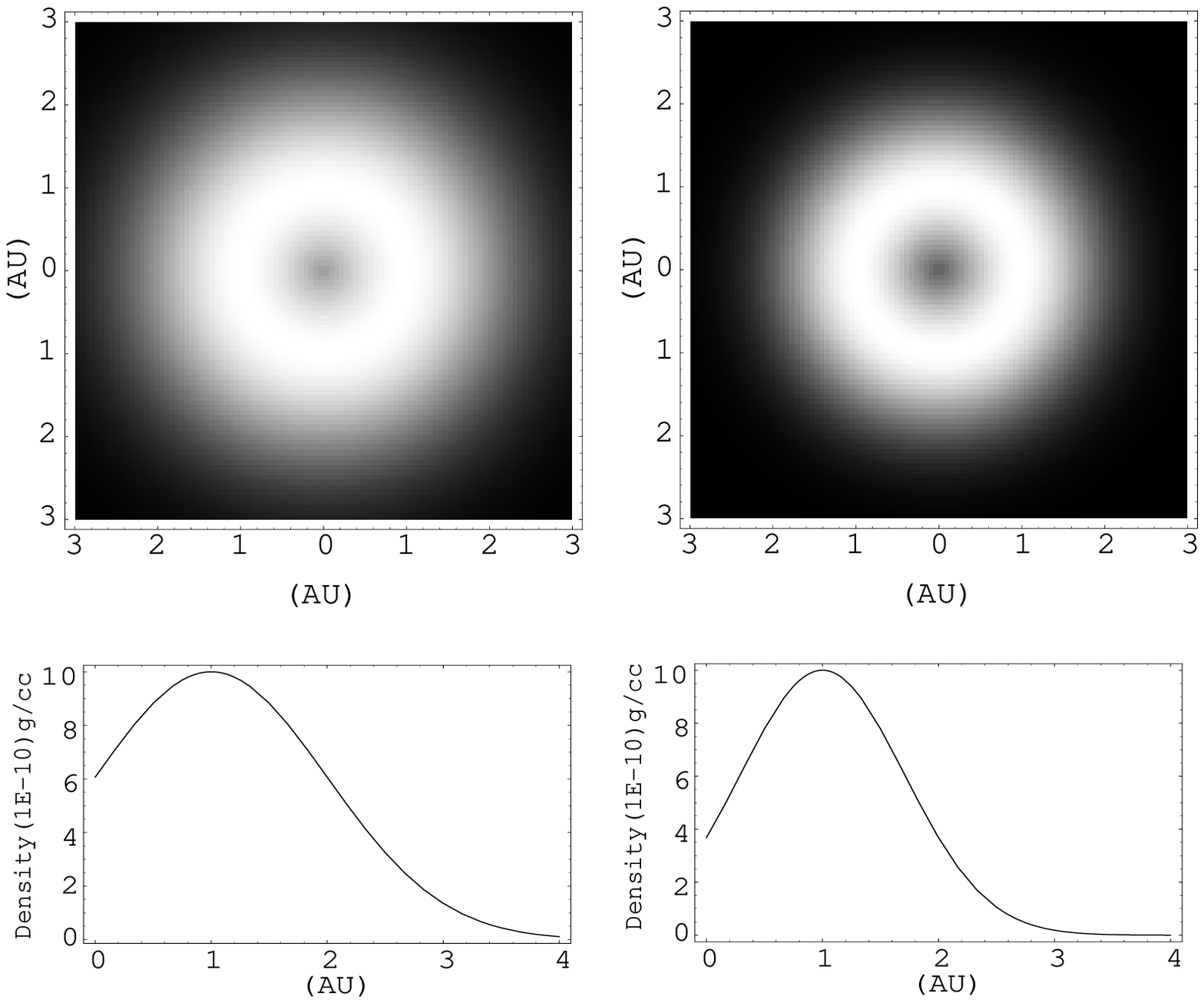}
\vskip -6in
\caption{ Graphs of the density of the gas for
$\alpha =0.5$ (left) and $\alpha =1$ (right)
in the disk midplane. \label{fig1}}
\end{figure}

\clearpage

\begin{figure}
\plotone{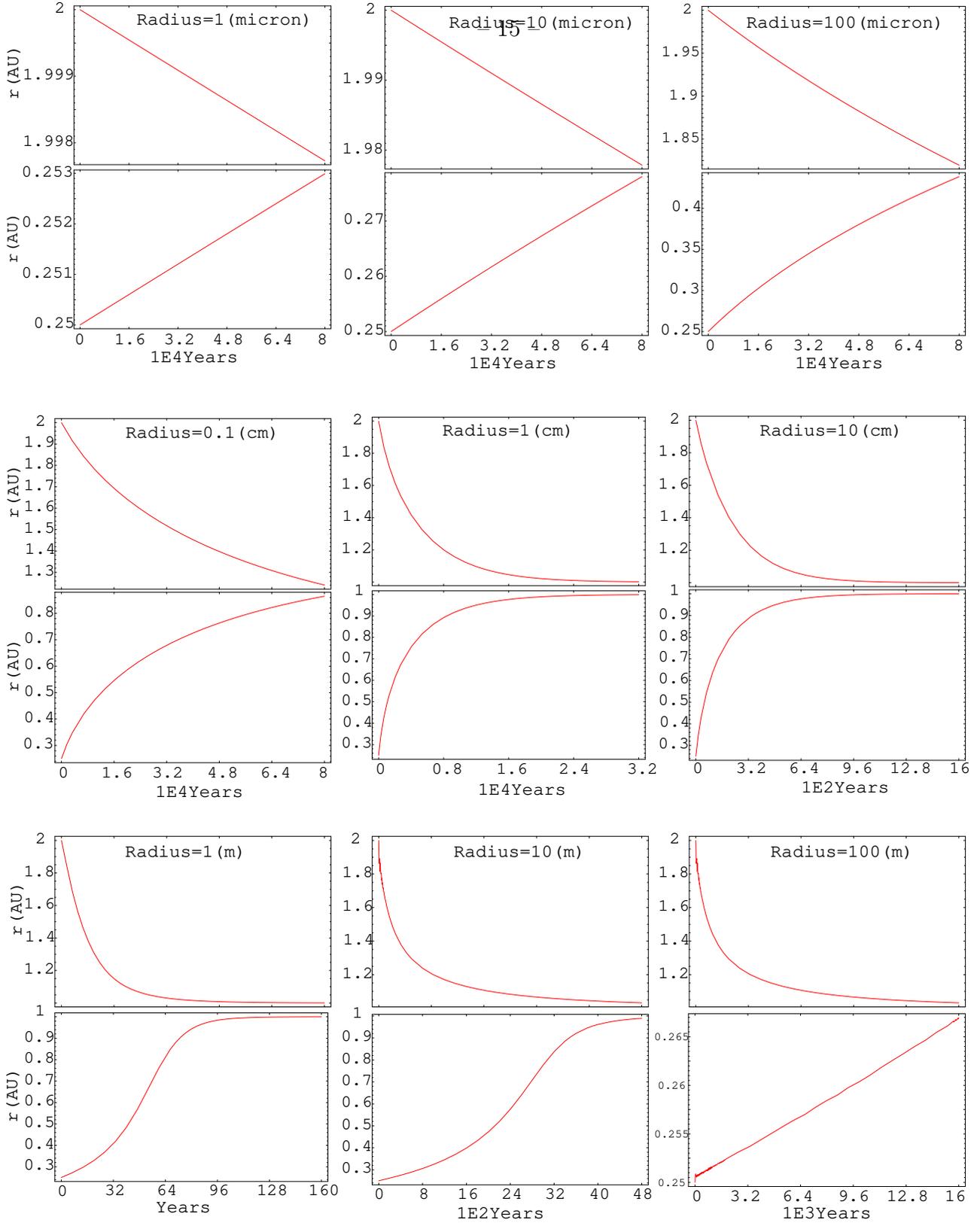}
\caption{Migration of solids with radii ranging from
1 micron to 100 m and solid particle densities equal to 2 g cm$^{-3}$. 
The disk is isothermal at 1000 K and its density is given by equation 
(17) with $\alpha=1$.  \label{fig2}}
\end{figure}

\clearpage

\begin{figure}
\plotone{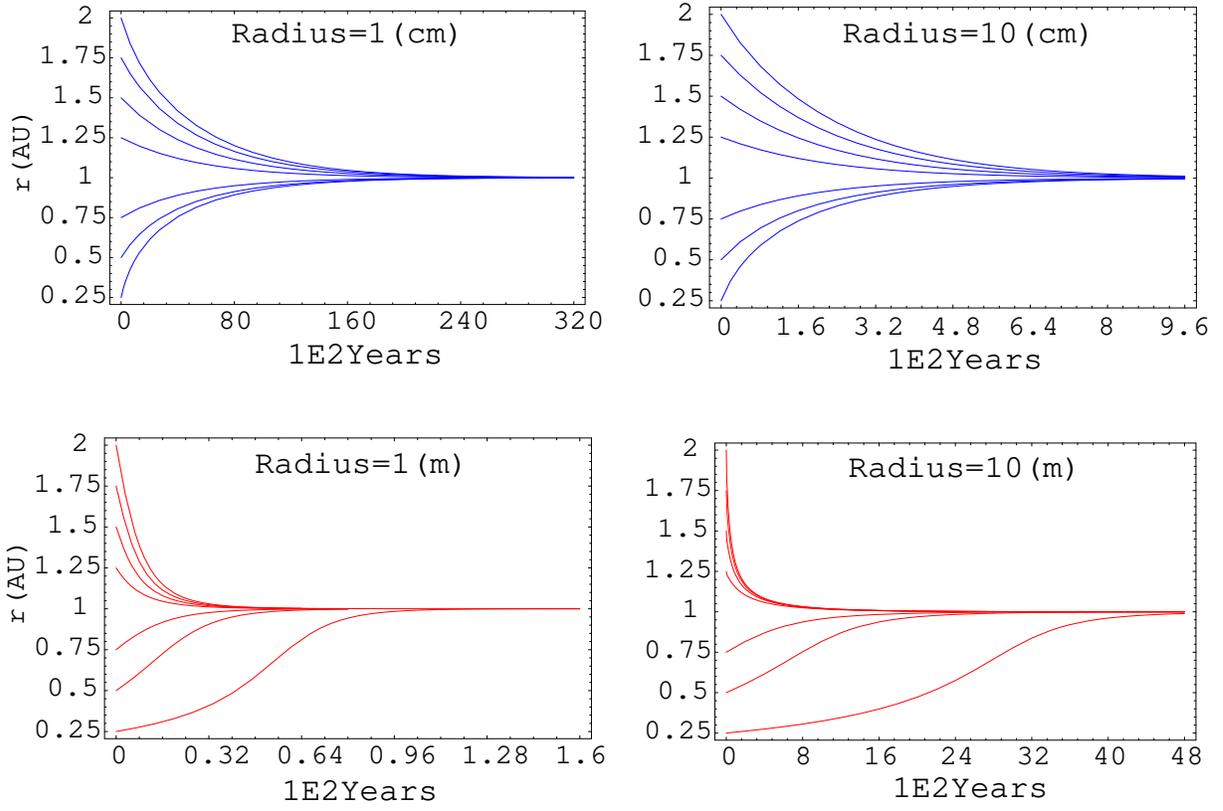}
\vskip -7in
\caption{Among the solids of Figure 1, the ones
with 10 cm to 10 m radii undergo rapid migrations. \label{fig3}}
\end{figure}

\clearpage

\begin{figure}
\plotone{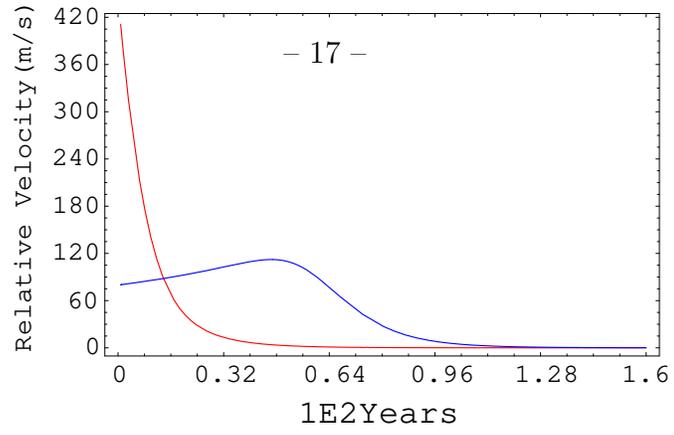}
\vskip -9in
\caption{Relative velocity of a 1 m-sized solid
while migrating inward from 1.75 AU (red) and outward from 0.25 AU 
(blue) to $r=1$ AU. \label{fig4}}
\end{figure}

\clearpage

\begin{figure}
\plotone{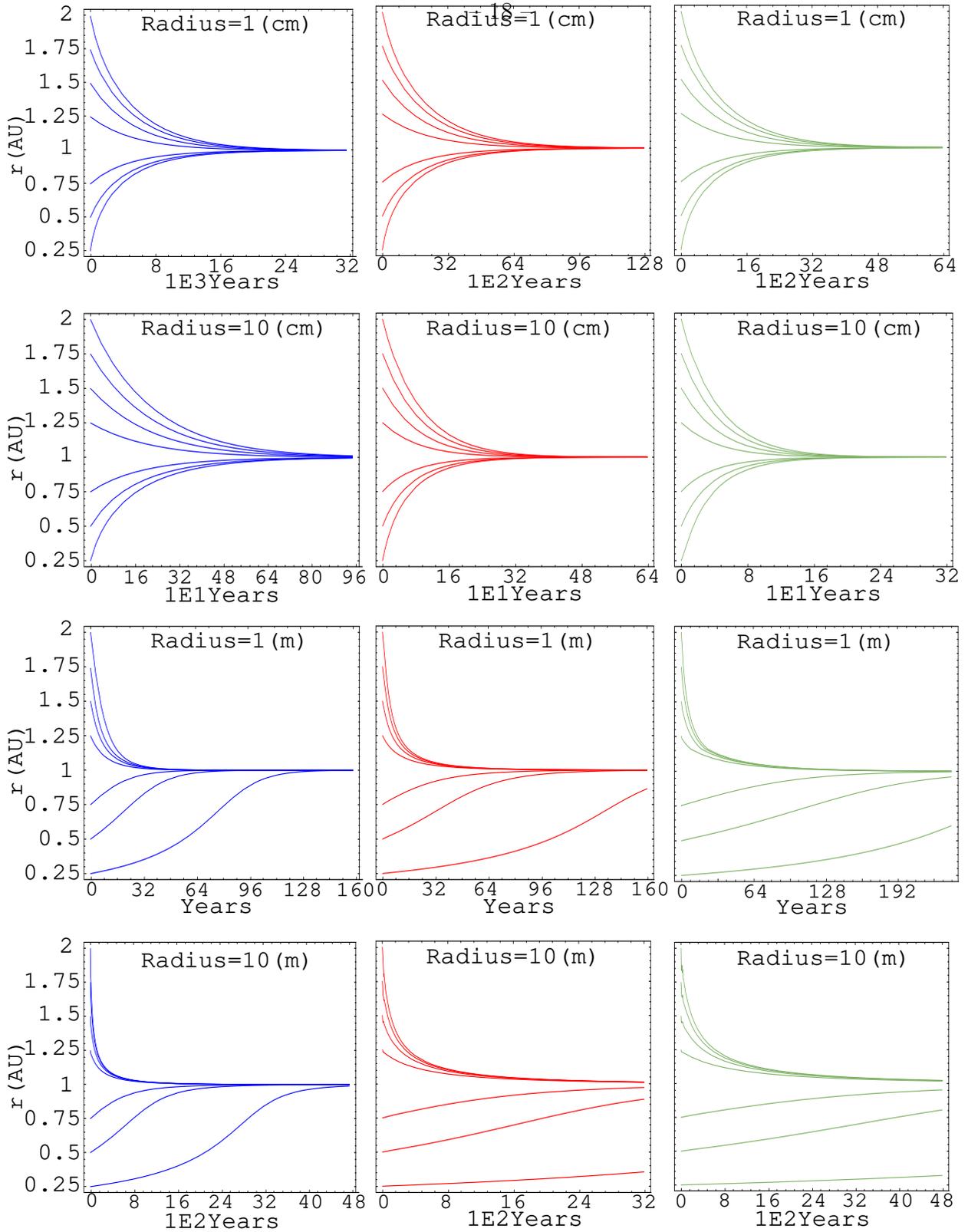}
\caption{ Migration of solids with densities
equal to 2 g cm$^{-3}$ (left column), 
5 g cm$^{-3}$ (middle column) and 10 g cm$^{-3}$
(right column). \label{fig5}}
\end{figure}

\clearpage

\begin{figure}
\plotone{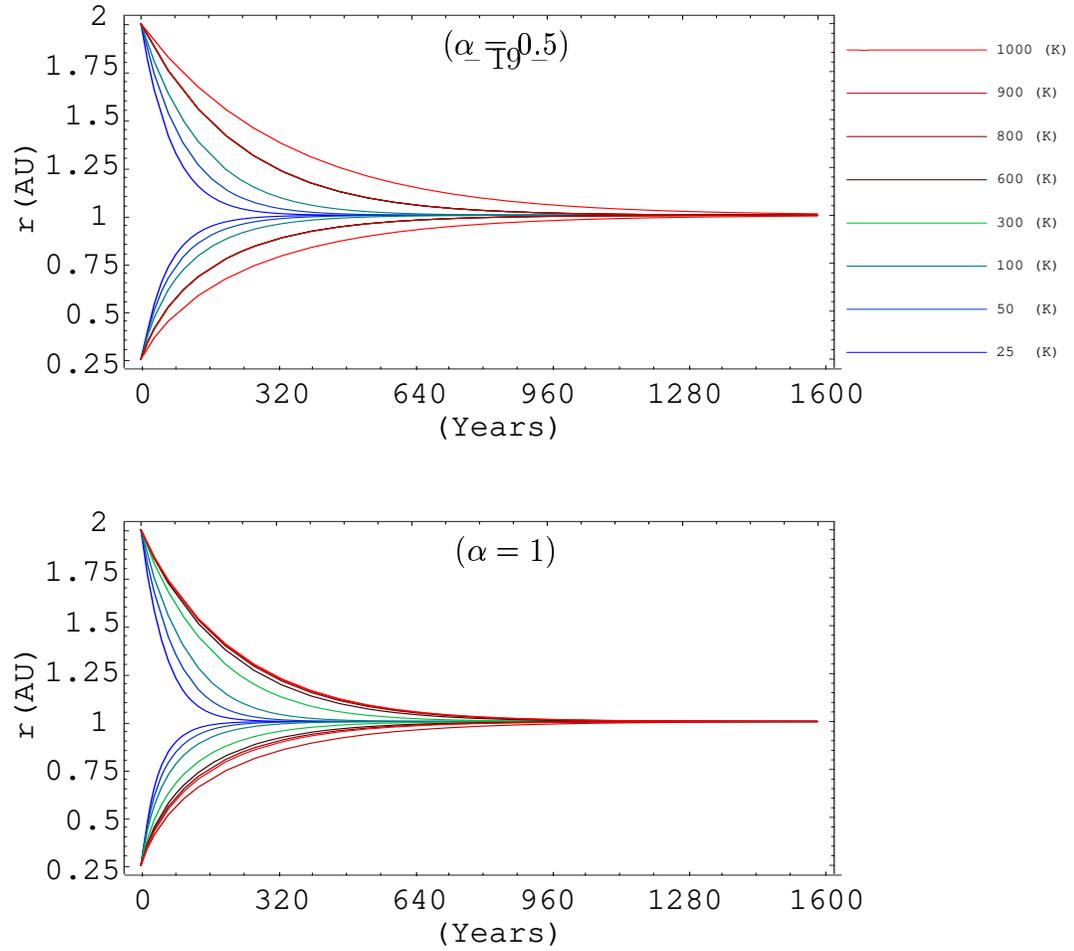}
\vskip -5.3in
\caption{ Migration of a solid at different
temperatures. The radius of the object is 10 cm and its density
is equal to 2 g cm$^{-3}$. As seen from this figure, the rate of 
migration decreases with increasing the temperature. \label{fig6}}
\end{figure}

\clearpage

\begin{figure}
\plotone{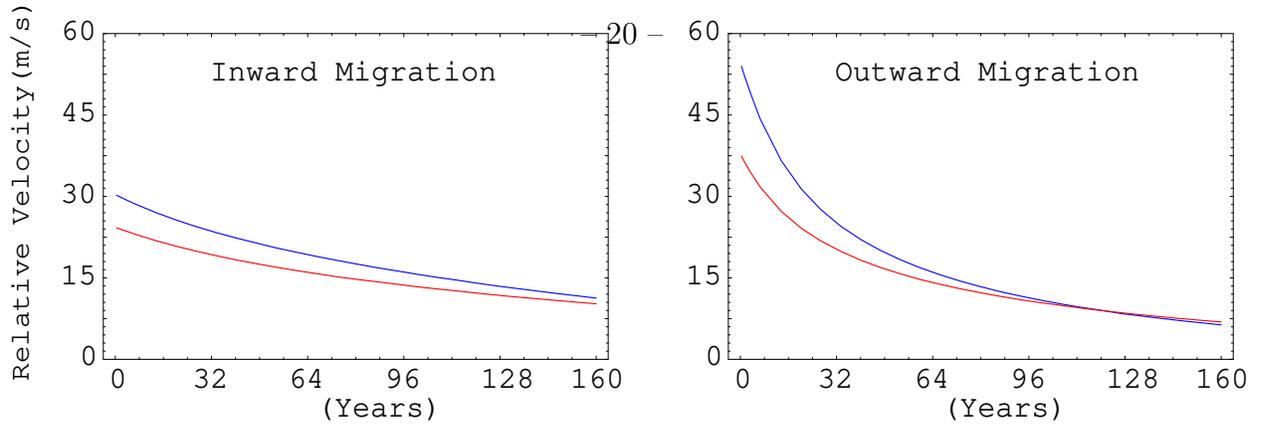}
\vskip -9in
\caption{ Relative velocity of a 10 cm-sized particle
with a density of 2 g cm$^{-3}$ migrating in a disk with a temperature
equal to 300 K (blue) and 1000 K (red). \label{fig7}}
\end{figure}

\end{document}